\documentclass[useAMS,usenatbib,letterpaper]{mn2e}
\usepackage{graphicx}
\usepackage{amsmath}
\usepackage{amssymb}

\title[AGN feedback triggered by gravitational collapse]{Investigating
  the properties of AGN feedback in hot atmospheres triggered by
  cooling-induced gravitational collapse}

\author[E.C.D. Pope, J.T. Mendel, S.S. Shabala] {Edward
  C.D. Pope$^{1}$\thanks{E-mail:ecdpope@uvic.ca}, J. Trevor
  Mendel$^{1}$, Stanislav S. Shabala$^{2}$\\$^{1}$Department of
  Physics \& Astronomy, University of Victoria, Victoria, BC, V8P 1A1,
  Canada\\$^{2}$School of Mathematics and Physics, University of
  Tasmania, Private Bag 37, Hobart, Tasmania 7001, Australia\\}

\begin{document}

\pagerange{\pageref{firstpage}--\pageref{lastpage} \pubyear{2011}}

\maketitle

\label{firstpage}

\begin{abstract}
Radiative cooling may plausibly cause hot gas in the centre of a
massive galaxy, or galaxy cluster, to become gravitationally
unstable. The subsequent collapse of this gas on a dynamical timescale
can provide an abundant source of fuel for AGN heating and star
formation. Thus, this mechanism provides a way to link the AGN
accretion rate to the global properties of an ambient cooling flow,
but without the implicit assumption that the accreted material must
have flowed onto the black hole from 10s of kiloparsecs away. It is
shown that a fuelling mechanism of this sort naturally leads to a
close balance between AGN heating and the radiative cooling rate of
the hot, X-ray emitting halo. Furthermore, AGN powered by
cooling-induced gravitational instability would exhibit characteristic
duty cycles ($\delta$) which are redolent of recent observational
findings: $\delta \propto L_{\rm X}/\sigma_{*}^{3}$, where $L_{\rm X}$
is the X-ray luminosity of the hot atmosphere, and $\sigma_{*}$ is the
central stellar velocity dispersion of the host galaxy. Combining this
result with well-known scaling relations, we deduce a duty cycle for
radio AGN in elliptical galaxies that is approximately $\propto
{M_{\rm BH}}^{1.5}$, where $M_{\rm BH}$ is the central black hole
mass. Outburst durations and Eddington ratios are also given. Based on 
the results of this study, we conclude that gravitational instability could 
provide an important mechanism for supplying fuel to AGN in massive 
galaxies and clusters, and warrants further investigation.
\end{abstract}


\section{Introduction}
The cooling time of X-ray emitting gas near the centres of many
massive galaxies and galaxy clusters is much shorter than the Hubble
time. In the absence of heat sources, significant quantities of the
gas would cool and form stars. However, X-ray spectroscopy has shown
that the rate at which gas cools to low temperatures is significantly
lower than first expected
\citep[e.g.][]{peterson01,tamura01,xu02,sak,peterson03,kaastra04,peterson06}
suggesting that the gas is somehow being reheated.

Based on both observational and theoretical evidence, it is generally
assumed that energy input by a central AGN is predominantly
responsible for reheating the gas. For example, elliptical galaxies
are commonly the hosts of powerful radio Active Galactic Nuclei
(AGN). These sources give rise to lobes of radio emission embedded in
the X-ray emitting gas which permeates massive galaxies and clusters
of galaxies \citep[e.g.][]{birzan,
  best05,dunn05,rafferty06,best08,stas}. Moreover, recent
observational studies of Brightest Cluster Galaxies (BCGs) suggest
that radio AGN activity is related to the thermal state of its
environment. Systems with short radiative cooling times, or a low
central entropy, are more likely to exhibit active star formation,
optical line-emission and jet-producing AGN
\citep[e.g.][]{burns,crawf,cav08,mittal,raff08}. This suggests that
AGN activity is part of a feedback loop that can prevent the ambient
hot gas from cooling, and is likely to have other important
consequences for its environment.

Building on the ideas of early work
\citep[e.g.][]{bintab,tucker,cfq,sr}, theoretical studies have drawn
attention to the potential, wide-ranging impact of AGN feedback. For
example, semi-analytic models of galaxy formation have demonstrated
that, in principle, AGN heating can both reheat cooling flows and
explain the exponential cutoff at the bright end of the galaxy
luminosity function \citep[e.g.][]{benson,croton05,bower06}, see also
\cite[][]{short}. More recently, AGN heating has been shown to be
crucial in shaping the X-ray luminosity-temperature relation of
massive galaxies \citep[e.g.][]{puchwein,bower08,pope09}.

Despite these findings, the fundamental details of AGN feedback remain
poorly constrained. Most significantly, there is no clear consensus on
how information about the thermal state of the X-ray emitting
atmosphere is transmitted to the black hole at the centre of the
galaxy, or cluster. Understanding this link is of great importance
because it probably facilitates the observationally-inferred long-term
balance between AGN heating and gas cooling, and almost certainly
governs the duty cycle of AGN activity -- that is, the fraction of
time an AGN is active.

Generally, AGN feedback is assumed to be powered by one of the
following: a) Bondi accretion of hot material in the vicinity of the
black hole \citep[e.g.][]{catt07,sijacki08,puchwein,fabj}; b) material
directly from the ambient cooling flow \citep[e.g.][]{pizz05,bower08,pope09,pizz10}. 
While there are plausible arguments for both, based on our current interpretation of
the available observations, there are also difficulties. For example,
the power output from Bondi accretion is unlikely to be sufficient to
balance the gas cooling rates in massive clusters \citep[e.g][]{soker06}. 
The biggest problem with models in category b) is finding a mechanism 
by which the central black hole receives information from the cooling flow 
about the thermal state of the ICM. For example, it is difficult to conceive of 
how material might flow all the way from the cluster cooling radius onto a 
central black hole, such that it prompts an AGN outburst on a useful 
timescale \citep[though see][]{pizz05}.

The model investigated here describes a mechanism that can reconcile the problem faced by models of type b). While \cite{pizz05,pizz10} focused on thermal instability of the ICM as a mechanism for delivering fuel to the AGN, we focus on the characteristics of AGN feedback that is triggered when the hot gas which resides near the centre of a galaxy becomes gravitationally unstable. That is, gas which was previously self-supporting against gravity is somehow destabilised and falls freely towards the galaxy centre, on a dynamical timescale, where it forms stars and fuels the
AGN \citep[e.g.][]{sr,fabian2009,king09b}. Gravitational instabilities
can be induced by merger events \citep[e.g.][]{sr} and the effect of
radiative cooling \citep[e.g.][]{birnboim03}. Of these two
possibilities, only instabilities induced by gas cooling can lead to
self-regulated AGN heating, but it is important to mention that
merger-driven AGN feedback will delay the onset of gravitational
instability induced by radiative cooling. The extent to which this
happens is unclear, but the effect may be more significant in group
and field environments \citep[e.g.][]{kaviraj}, than in clusters.

Perhaps importantly, AGN heating driven by cooling-induced
gravitational instabilities has the potential to be periodic, as
explained by the following argument. In massive galaxies and clusters
there is a continual inflow of material from the large-scale
environment, due to the effects of the ambient cooling flow. As the
mass of material in the hot atmosphere near the centre of a galaxy
builds up, it can become gravitationally unstable, meaning that some
fraction of the gas will collapse and flow on a dynamical timescale to
the centre of the gravitational potential, thereby fuelling the
AGN. Once this fuel has been consumed, the AGN will be starved of new
material until the nearby hot gas once again becomes gravitationally
unstable. A second instability will commence after the external inflow
has delivered a sufficient quantity of new material. Thus, for this
mode of fuelling, the typical duration of an AGN outburst is largely
governed by the local dynamical timescale, while the time between the
onset of successive outbursts is controlled by the ambient cooling
flow. This is notable because observations indicate that AGN are only
active for a fraction of time determined by their environment and host
galaxy properties \citep[e.g.][]{best05,stas}. In addition, as
demonstrated in \cite{pope11}, periodic heating can supply the energy
required to balance gas cooling, with the minimum effort. Thus,
periodic AGN activity appears to be an energetically favourable
heating strategy. Therefore, starting from the assumption that AGN
fuelling is attributable to gravitational instabilities induced by gas
cooling, we derive AGN duty cycles, outburst durations, star formation rates and discuss implications for numerical
simulations.

The outline of the article is as follows. In Section 2 we present the
main model and place constraints on the phenomenological
parameters. In Section 3 we show the expected duty cycles and outburst
durations that would be associated with AGN
heating that is driven by cooling-induced gravitational
instability. The implications are discussed in Section 4, and the
results are summarised in Section 5.

\section{Outline of the model}

In this investigation, we study the properties of AGN feedback that is
driven by the gravitational collapse of hot gas in the vicinity of the
supermassive black hole. Therefore, we must first describe the
physical conditions of the gas in massive galaxies and clusters that
can justifiably lead to gravitational instability. Following this we
define the characteristic timescales of the system and the AGN power
output expected from this mode of fuelling. These quantities are then
used to determine the observable features predicted by this scenario.

\subsection{The onset of gravitational instability}

A self-gravitating sphere of gas will become gravitationally unstable
when the sound-crossing time is greater than the gravitational
free-fall time. This is the well-known Jeans criterion. Strictly
speaking, the criterion only applies in the limit that the polytropic
index of the gas -- $\gamma_{\rm eff}$, as defined in equation
(\ref{eq:1}) -- is also lower than some critical value, $\gamma_{\rm
  crit} \approx 1.2$, \citep[e.g. see][and references
  therein]{birnboim03}. Since the \emph{actual} adiabatic index of an
ideal, non-relativistic monatomic gas is $\gamma = 5/3$, a gas of this
type can only become gravitationally unstable if its behaviour is
modified by heating, cooling and work. For example, if 
a parcel of gas is heated by an amount $\Delta Q$, and then does an
equal amount of work on its environment (i.e. $\Delta W = \Delta Q$),
the gas is defined as isothermal, since it acts to maintain a constant
temperature. In this case, the polytropic index of the gas must be
$\gamma_{\rm eff} = 1$, by definition.

In the case of the hot, X-ray emitting gas that resides in massive
galaxies and galaxy clusters, we envisage a central gas mass that is
embedded in a pressurised ambient medium. Consequently, the scenario
is somewhat reminiscent of the isothermal sphere which is prone to
gravitational instability if its mass exceeds the Bonnor-Ebert
limit. However, it is important to note that \emph{in this study we do
  not make any assumptions about the density distribution of the hot
  gas}. 
  
Observations clearly indicate that this gas must be subject to
radiative cooling, heating from AGN and stars, \emph{and} the work
done by its surroundings. Therefore, it seems extremely likely that
$\gamma_{\rm eff} \neq 5/3$ for the X-ray emitting gas. However, it is
unclear exactly what values the polytropic index might take because
the processes acting on the gas are highly uncertain. Nevertheless,
consider the behaviour of a gas mass near the centre of a massive
galaxy cluster. As the gas radiates energy it loses pressure support;
the weight of the overlaying gas then does work and compresses
it. The contraction can only persist if the compressional heating 
resulting from the inflow is radiated away on a timescale that is shorter than the 
dynamical timescale, i.e. $t_{\rm cool} < t_{\rm dyn}$. This is perfectly possible because 
a gas parcel of temperature $T$ and number density $n$, radiating via 
thermal bremsstrahlung, has a cooling time that is $\propto T^{1/2}/n$. As the gas 
cools, the temperature will fall, and the density will rise, so that the cooling time 
shortens. It is this condition ($t_{\rm cool} < t_{\rm dyn}$) that explains observations
which show that the temperature of the gas remains lower near the centre of the 
cluster than further out. Indeed, it is precisely this argument that predicts a 
cooling catastrophe, and the need for AGN feedback. However, as shown below, 
the inflow condition also predicts the onset of a cooling-induced 
gravitational instability which provides a potentially informative description of 
AGN fuelling in hot atmospheres.

Thermodynamically, the mass of gas described above behaves like a parcel that 
cools down (rather than heats up) when compressed. The appropriate polytropic 
relation linking the temperature, $T$, and volume, $V$, of the gas parcel is 
$T V^{\gamma_{\rm eff}-1} = {\rm constant}$. Thus, the hot gas in the centres 
of massive galaxies and clusters acts as if $0 < \gamma_{\rm eff} < 1$. Since this is less 
than all plausible values of the critical polytropic index, the hot gas is likely to be 
susceptible to gravitational instability \footnote{The gas exhibits characteristics which 
are somewhere between isothermal ($\gamma_{\rm eff} = 1$) and isobaric 
($\gamma_{\rm eff} = 0$) states.}.

For completeness, we also present a brief derivation to illustrate the
general phenomena that influence the polytropic index of a gas. This
derivation closely follows the work of \cite{birnboim03} who define
the polytropic index, $\gamma_{\rm eff}$, of a Lagrangian fluid
element in terms of the logarithmic time derivative of its pressure,
$P$, and density, $\rho$, such that
\begin{equation}\label{eq:1}
\gamma_{\rm eff} \equiv \frac{{\rm d}\ln P /{\rm d}t}{{\rm d}\ln \rho
  /{\rm d}t}.
\end{equation}
Accounting for gas cooling, the polytropic index can be expressed as
\begin{equation}
\gamma_{\rm eff} = \gamma - \frac{n^{2}\Lambda(T)}{\dot{\rho}e},
\end{equation}
where $\gamma$ is the actual adiabatic index of the gas, $\dot{\rho}$
is the rate of change of gas density and $e$ is the internal energy
per unit mass. As usual, $n$ is the number density of the gas, $T$ is
the gas temperature and $\Lambda(T)$ is the cooling function.

\subsection{Characteristic timescales and flow rates}

To calculate the local dynamical timescale, we refer back to the
definition of the Jeans instability criterion. A self-gravitating gas mass will
become gravitationally unstable when its mass exceeds some critical
value that is comparable to the Jeans/Bonnor-Ebert masses. In the
centre of a galaxy, the self-gravity of the gas will become important
when its characteristic velocity dispersion is comparable to the
central stellar velocity dispersion, $\sigma_{*}$. Thus, if a gas mass
$M$ with radius $R$ located at the centre of a galaxy becomes
gravitationally unstable, it will collapse on the dynamical timescale,
$t_{\rm dyn} = R/\sigma_{*}$. The average mass flow rate during this
time will be
\begin{equation}
\dot{M} = {\frac{\Delta M}{\Delta t}} \approx \frac{M}{t_{\rm dyn}} =
\frac{\sigma_{*}M}{R}
\end{equation}
We note that, for a self-gravitating cloud, the characteristic
velocity dispersion of the constituent particles is related to the
gravitational potential energy by the virial theorem: $\alpha G M / R
= \sigma^{2}$, where $G$ is Newton's gravitational constant. In this
description the numerical constant $\alpha \sim 1$ depends on the
density distribution of the gas. Thus, the mass flow rate can be
expressed as
\begin{equation} \label{eq:dotm}
\dot{M} = \frac{\sigma_{*}^{3}}{\alpha G} \sim 1000
\bigg(\frac{\sigma_{*}}{200\,{\rm km\,s^{-1}}}\bigg)^{3}\,{\rm
  M_{\odot}\,yr^{-1}}.
\end{equation}
This is the well-known form of the dynamical mass flow rate
\citep[c.f.][]{king09b}, modified slightly to account for arbitrary
density distributions. From this it is straightforward to show that
the duration of the collapse will be $M/\dot{M} = t_{\rm dyn}$.

Following the collapse, mass will flow towards the centre of the
galaxy's gravitational potential where a fraction is likely to be
accreted by the supermassive black hole, and the remainder 
presumably forms stars nearby or is expelled from the galaxy 
as a result of feedback.

As described in the introduction, the fuelling rate provided by
gravitational instability can be periodic, for the following
reason. The postulated gas cloud near the galaxy centre acts as a
reservoir; this reservoir is depleted when the cloud collapses, but is
replenished by the inflow of new material from the larger-scale
environment, which forms a new cloud. When it reaches the critical
mass, the second cloud will also become gravitationally unstable and
collapse, and so on. 

The provenance of the material that replenishes 
the collapsing gas cloud is not clear, except that it must come from the 
ambient hot atmosphere. The simplest possibility is that the cloud is built 
up during two phases: 1) an initial collapse of material from slightly further out 
in the hot atmosphere; 2) the subsequent slow inflow of additional material due to the
ambient cooling flow. If this is the case, the first phase of growth of the new cloud 
will occur on a similar dynamical timescale to the gravitational
collapse of the previous cloud. However, at these early times, the new cloud is likely 
to be hotter and of lower mass than the previous cloud, ensuring that it can be 
gravitationally stable. More precisely, as long as $\gamma_{\rm eff}  < \gamma_{\rm crit } \approx 1.2$, 
and the cloud mass is below the critical mass, the cloud will not collapse. The second 
phase of growth will occur on the timescale required for the cooling 
flow to build up the cloud mass to its gravitationally unstable limit, at which point a 
collapse will be initiated. Using this argument, we can estimate the average time between 
gravitational collapses, as shown below.

Overall, the local mass inflow rate, $\dot{M}_{\rm ext}$, from the cooling flow will 
be a slowly varying quantity governed by the difference 
between the time-averaged heating and cooling rates. Then, in the limit that 
the majority of the cloud mass is built up during this phase, the characteristic 
time between the triggering of successive gravitational collapses must tend to
\begin{equation} \label{eq:tau}
\tau = \frac{M}{\dot{M}_{\rm ext}}.
\end{equation}
Assuming that the AGN is only fuelled while the cloud collapses,
it will be active for a fraction of time $\delta = t_{\rm dyn}/\tau$, known as the duty cycle. By combining equations (\ref{eq:dotm}) and
(\ref{eq:tau}), we obtain a simple functional form for the AGN duty
cycle without having to explicitly calculate the AGN heating rate
\begin{equation} \label{eq:duty}
\delta = \frac{t_{\rm dyn}}{\tau} = \frac{\dot{M}_{\rm ext} R}{M
  \sigma_{*}} = \frac{\alpha G \dot{M}_{\rm ext}}{\sigma_{*}^{3}}.
\end{equation}
Thus, in the present model, the AGN duty cycle can be
straightforwardly related to the local gravitational potential,
through $\sigma_{*}$, and the external environment through
$\dot{M}_{\rm ext}$. This is potentially significant, because
obervations \citep[e.g.][]{best05,best07,stas} indicate that the radio
AGN duty cycle is heavily influenced by both local and environmental
effects. Below, we argue that $\dot{M}_{\rm ext}$ is probably closely
related to the mass inflow associated with the ambient cooling flow.

\subsection{AGN heating rates}

If a fraction, $\beta$, of the inflowing mass rate, $\dot{M}$, reaches
the black hole, the accretion power output will be
\begin{equation}\label{eq:h}
H = \eta \beta \dot{M} c^{2} = \eta \frac{\beta \sigma_{*}^{3}}{\alpha
  G} c^{2} \approx 10^{47}
\bigg(\frac{\beta/\alpha}{10^{-3}}\bigg)\bigg(\frac{\sigma_{*}}{200\,{\rm
    km\,s^{-1}}}\bigg)^{3}\,{\rm erg\,s^{-1}},
\end{equation}
where the assumed accretion efficiency is $\eta \approx 0.1$. The
characteristic value of $\beta/\alpha = 10^{-3}$ is motivated below,
and yields favourable comparisons with the observationally-inferred
radio AGN duty cycle.

Clearly, the AGN energy injection rates associated with this fuelling
mechanism can be very large. Such values are considerably larger than
would be expected to arise from the accretion of nearby hot gas and,
if observed in a real system, would more commonly be associated with
merger-driven fuelling events. However, the reasoning above indicates
that such an interpretation is not necessarily exclusive -- high AGN
fuelling rates can also be a consequence of gravitational instability
resulting from gas cooling. In addition, it is important to remember
that values estimated from equation (\ref{eq:h}) represent the
instantaneous heating rate -- the time-averaged values are much more
modest.

By definition, the time-averaged AGN power output is written $\bar{H}
\equiv \delta H$, which can be expanded using equations
(\ref{eq:duty}) and (\ref{eq:h}) to give
\begin{equation}\label{eq:barh}
\bar{H} = \delta H = \eta \beta \dot{M}_{\rm ext} c^{2}.
\end{equation}
Equation (\ref{eq:barh}) shows that the time-averaged AGN heating rate
depends on the properties of the large scale environment, through
$\dot{M}_{\rm ext}$. From this we also conclude that $\dot{M}_{\rm
  ext}$ will evolve until the time-averaged heating rate closely
matches the ambient cooling rate of the gas. In the limit that the
time-averaged AGN heating does balance gas cooling, $\bar{H} = L_{\rm
  X}$, equation (\ref{eq:barh}) shows that
\begin{equation}\label{eq:mext}
\dot{M}_{\rm ext} = \frac{L_{\rm X}}{\eta \beta c^{2}} \approx 175
\bigg(\frac{L_{\rm X}}{10^{45}\,{\rm
    erg\,s^{-1}}}\bigg)\bigg(\frac{\beta}{10^{-3}}\bigg)^{-1}\,{\rm
  M_\odot\,yr^{-1}},
\end{equation}
where again we have assumed $\eta \approx 0.1$ and $\beta/\alpha \sim
10^{-3}$, with $\alpha \sim 1$. Below, we motivate the constraints on
$\beta$ by considering the Eddington ratio of the AGN outbursts
fuelled by gravitational collapse.

\subsection{The Eddington ratio}

The Eddington ratios predicted by this model provide another method for comparison 
with numerical simulations and observations. Importantly, they also provide a way to check 
the self-consistency of the model, as outlined below.

Broadly speaking, the form of energetic output from an AGN can be
predicted from its Eddington ratio, defined by
\begin{equation}
R_{\rm Edd} \equiv \frac{\dot{m}}{\dot{M}_{\rm Edd}},
\end{equation}
where $\dot{m}$ is the black hole accretion rate. $\dot{M}_{\rm Edd}$
is the Eddington limited accretion rate determined by the balance
between gravity and radiation pressure
\begin{equation}\label{eq:edd}
\dot{M}_{\rm Edd} = \frac{L_{\rm Edd}}{\eta c^{2}} = \frac{4\pi G
  m_{\rm p}M_{\rm BH}}{\eta c \sigma_{\rm T}},
\end{equation}
where $L_{\rm Edd}$ is the Eddington luminosity, $M_{\rm BH}$ is the
black hole mass, $\sigma_{\rm T}$ is the Thompson cross-section, $m_{\rm
  p}$ is the proton mass and the other symbols are as previously
defined.

By analogy with stellar mass black holes in X-ray binaries
\citep[e.g.][]{elmar}, accretion at rates less than $\sim$3\% of the
Eddington limit is radiatively inefficient so that the majority of the
power output is in the form of kinetic-energy-dominated outflows of
relativistic particles which are prominent radio synchrotron
emitters. Outflows of this type are thought to couple strongly to the ambient gas. 
Conversely, accretion above this critical rate is radiatively efficient, meaning
that the power output is predominantly in the form of photons, rather
than kinetic outflows of particles. In this limit, radio jets may still be observed, but the efficiency
of jet production is much lower than in the radiatively inefficient
regime \citep[e.g.][]{macc}. Furthermore, in the radiatively efficient mode, it has
been argued that only $\sim$5\% of the accretion power is available to
heat the surrounding gas \citep[e.g.][]{sijacki06}, see \cite{king09b}
for a possible explanation. 

We note that the comparison with X-ray binaries is not exact, since it does not account 
for the existence of radio-loud quasars which produce powerful kinetic outflows at high 
Eddington ratios. However, as explained below, we focus on low accretion rate objects 
so the distinction does not matter for the purposes of this model.

To calculate the Eddington ratios, we use the well-known
relation between the black hole mass and the stellar velocity
dispersion: $M_{\rm BH} \approx 1.3 \times 10^{8}(\sigma_{*}/200\,{\rm
  km\,s^{-1}})^{4}\,{\rm M_\odot}$, where $\sigma_{*}$ is the central
stellar velocity dispersion \citep[e.g.][]{tremaine}. Substituting for
the black hole mass into equation (\ref{eq:edd}) yields the Eddington
ratio for an AGN fuelled by gravitational collapse
\begin{eqnarray}\label{eq:edf}
R_{\rm Edd} \approx 0.01
\bigg(\frac{\beta/\alpha}{10^{-3}}\bigg)\bigg(\frac{\sigma_{*}}{200\,{\rm
    km\,s^{-1}}}\bigg)^{-1},
\end{eqnarray}
where the assumed accretion efficiency is $\eta \approx
0.1$. Therefore, if $\beta/\alpha \lesssim 10^{-3}$, accretion
following a gravitational collapse can be radiatively inefficient and
produce kinetic outflows. Since $\alpha \sim 1$, this
implies that $\beta \lesssim 10^{-3}$ is required for kinetic outflow
production. For larger values of $\beta/\alpha$ the accretion will be
radiatively efficient. 

The value of $\beta$ is depends on the processes that govern how 
gas travels from kiloparsec scales onto the black hole. Since these processes 
are highly uncertain, we proceed by assigning $\beta$ a single, empirical value 
that encapsulates the accretion physics in the full range of AGN environments. 
While this is a simplification, it ensures that the results are as transparent as 
possible. As indicated in equation (\ref{eq:edf}), we find that $\beta \sim 10^{-3}$ 
permits a good agreement between the model and radio observations 
\citep[][]{best05,best07,stas}. Reassuringly, equation (\ref{eq:edf}) also shows that $\beta \sim 10^{-3}$ will lead 
to low Eddington ratios and, therefore, kinetic outflows which produce radio emission.

\section{Results}

From the assumptions and derivations outlined above, we present the
expected duty cycles of radio AGN fuelled by gravitationally
destablised gas, the corresponding duration of individual outbursts,
and the associated heating rates resulting from shocks generated by
the AGN outflow.

\subsection{AGN duty cycle}

In this section, we give a more general derivation for the AGN duty
cycle. For a constant pressure cooling flow in which external heating
is not important, the classical mass flow rate, gas temperature, and
X-ray luminosity are related by
\begin{equation}
L_{\rm X} = \frac{\gamma}{\gamma-1}\dot{M}_{\rm clas}\frac{k_{\rm
    B}T}{\mu m_{\rm p}},
\end{equation}
where $k_{\rm B}$ is Boltzmann's constant and $\mu m_{\rm p}$ is the
mean mass per particle.

More generally, the net inflow of mass is determined by the difference
between the time-averaged cooling and heating rates. Observationally,
this mass flow rate is estimated by fitting models to the X-ray
spectra. Thus, we refer to the mass flow rate by its observational
name, $\dot{M}_{\rm spec}$, but calculate it from the following
\begin{equation}\label{eq:bal}
L_{\rm X} - \bar{H} = \frac{\gamma}{\gamma-1}\dot{M}_{\rm
  spec}\frac{k_{\rm B}T}{\mu m_{\rm p}}.
\end{equation}
Rearranging equation (\ref{eq:bal}) for the duty cycle, using equation
(\ref{eq:h}), gives
\begin{equation}
\delta = \frac{\alpha G L_{\rm X}}{\eta \beta c^{2}
  \sigma_{*}^{3}}\bigg(1 - \frac{\dot{M}_{\rm spec}}{\dot{M}_{\rm
    clas}}\bigg).
\end{equation}
Consequently, in the limit that AGN heating exactly balances gas
cooling, $\dot{M}_{\rm spec}/\dot{M}_{\rm clas} \rightarrow 0$, the
duty cycle simplifies to
\begin{equation}
\delta \rightarrow \frac{\alpha G L_{\rm X}}{\eta \beta c^{2}
  \sigma_{*}^{3}} \propto \frac{L_{\rm X}}{\sigma_{*}^{3}}.
\end{equation}
In terms of scaled quantities, the duty cycle can be expressed as
\begin{equation}\label{eq:duty2}
\delta \approx 0.1 \bigg(\frac{\beta/\alpha}{10^{-3}}\bigg)^{-1}
\bigg(\frac{L_{\rm X}}{10^{45}\,{\rm
    erg\,s^{-1}}}\bigg)\bigg(\frac{\sigma_{*}}{200\,{\rm
    km\,s^{-1}}}\bigg)^{-3}.
\end{equation}
Equation (\ref{eq:duty2}) shows that AGN fuelled by the gravitational
collapse of gas in more X-ray luminous clusters should exhibit larger
duty cycles. That is, those AGN should be active for a greater
proportion of time, as is inferred from observations
\citep[][]{best08}. However, the precise scaling of the AGN duty cycle
in clusters is difficult to determine because it is not clear exactly
how the X-ray luminosity of the gas, and the central stellar velocity
dispersion are related in clusters. Nevertheless, surveys indicate
that the X-ray luminosity of a cluster scales as $L_{\rm X} \propto
\sigma_{\rm c}^{n}$, where $\sigma_{\rm c}$ is the velocity dispersion
of the cluster potential, and that $n \sim 4-5$
\citep[e.g.][]{mahdavi}. Substituting the cluster $L_{\rm
  X}-\sigma_{\rm c}$ relation into equation (\ref{eq:duty2}), gives
$\delta \propto \sigma_{\rm c}^{n}/\sigma_{*}^{3}$. Therefore, if more
massive BCGs are found in more massive clusters, the AGN duty cycle
will tend to increase slowly with $\sigma_{*}$, which would be in
qualitative agreement with observations
\citep[e.g.][]{best08}. Furthermore, assuming $\beta/\alpha \sim 10^{-3}$,
the normalisation of the duty cycle is entirely consistent with the
observational results presented by \cite{best08}.

In contrast, the scaling relations of field elliptical galaxies yield
a much simpler scaling of $\delta$ with $\sigma_{\rm *}$. Under the
assumption of an isothermal gravitational potential, the velocity
dispersion is constant and independent of radius. Again, surveys
indicate that the X-ray luminosity scales as $L_{\rm X} \propto
\sigma_{*}^{m}$, where $m \sim 8-10$
\citep[e.g.][]{osullivan03,mahdavi}. Using this fact, the AGN duty
cycle would scale as $\delta \propto \sigma_{*}^{m-3} \sim
\sigma_{*}^{6}$. Since supermassive black hole mass scales as $M_{\rm
  BH} \propto \sigma_{*}^{4}$, the duty cycle can then be said to
increase as $\delta \propto M_{\rm BH}^{1.5}$, which is consistent
with observational findings \citep[][]{best05,stas}. In addition, for a
massive galaxy with an X-ray luminosity of $10^{41}\,{\rm
  erg\,s^{-1}}$, and $\sigma_{*} \sim 200 \,{\rm km\,s^{-1}}$ the duty
cycle would be $\delta \approx 10^{-4}$ (assuming $\beta/\alpha \sim
10^{-3}$), which is also consistent with
\cite{best05} and \cite{stas}. Consequently, both the scaling and
normalisation of the duty cycle are largely consistent with
observations of radio AGN in field elliptical galaxies and BCGs.

\subsection{Number and duration of outbursts}

The AGN duty cycle is a useful quantity which provides information
about how frequently an AGN is triggered in order for heating to
balance the effects of radiative cooling. With additional assumptions,
the duty cycle also provides some information about the duration of an
individual heating event, $t_{\rm dyn}$, and its environmental
dependence. If the time between the onset of successive AGN outbursts
is $\tau$, the number of heating events which occur during a galaxy
lifetime, $t_{\rm age}$, must be $N = t_{\rm age}/\tau = \delta
(t_{\rm age}/t_{\rm dyn})$. Using the previous results, the expected
number of AGN outbursts can be written
\begin{eqnarray}\label{eq:N}
N \approx 10\bigg(\frac{\beta/\alpha}{10^{-3}}\bigg)^{-1}
\bigg(\frac{t_{\rm dyn}/t_{\rm
    age}}{0.01}\bigg)^{-1}\bigg(\frac{L_{\rm X}}{10^{45}\,{\rm
    erg\,s^{-1}}}\bigg) \\ \nonumber \times \bigg(\frac{\sigma_{*}}{200 \,{\rm
    km\,s^{-1}}}\bigg)^{-3}.
\end{eqnarray}
By rearranging equation (\ref{eq:N}), the duration of an outburst can
be expressed in the form $t_{\rm dyn} = \delta t_{\rm age}/N$. Using
equation (\ref{eq:duty2}) for a cluster with $L_{\rm X} \sim
10^{45}\,{\rm erg\,s^{-1}} $ and $\sigma_{*} \sim 200 \,{\rm
  km\,s^{-1}}$, the typical outburst duration must be $t_{\rm dyn}
\sim 0.01 t_{\rm age}$. Taking $t_{\rm age} \sim 5\,{\rm Gyr}$ implies
$t_{\rm dyn} \sim 50 \,{\rm Myr}$, which is sufficient to explain the
features AGN-blown bubbles observed in many clusters
\cite[e.g.][]{birzan,dunn05}.

The gravitational instability model of AGN fuelling also offers an
explanation for the unexpectedly large number of compact radio sources
\citep[e.g.][]{odea,stas} that differs from the accretion disk
variability model proposed by \cite{czerny09}. As shown by
\cite{alex00}, the maximum stable length of an AGN jet propagating
through an atmosphere depends on its power and the ambient
density. For example, at constant jet power, a higher ambient density
leads to a shorter stable jet length. Then, since the inflow of
material due to gravitational instability will significantly enhance
the density in the vicinity of the black hole, this enhanced density
may also plausibly confine and disrupt the resultant AGN jet on
kiloparsec scales. If this is correct, the jet can only propagate
further outwards once the gas has been sufficiently depleted by
continued accretion and star formation.

\section{Discussion}

The model investigated above exhibits several features that are
compatible with key observational characteristics of AGN feedback in
both field elliptical galaxies and BCGs. Below we discuss some
additional implications of the model which may be significant, but are
harder to quantify.

\subsection{Star formation in the host galaxy}

Since only a very small fraction of the inflowing material reaches the
black hole, the vast majority must either form stars, or be dragged
out of the galaxy by the AGN outflow itself. As shown in
\cite{popeetal} the mass of ambient material transported by an
AGN-blown bubble is approximately equal the mass of gas initially
displaced by the bubble. In principle, this process can be extremely
effective at removing material from the galaxy, dramatically reducing
the quantity of gas available for forming stars. As a result, it is
difficult to quantify star formation rates resulting from
cooling-induced gravitational instability, because it is unclear how
much of the collapsed gas will be retained by the galaxy, and for how
long. Our current best estimate is that the mass of material available
for forming stars must fall between two well-defined limits. The upper
limit is the case in which all of the collapsed gas mass goes into
forming stars; the lower limit is the case in which the vast majority
of the collapsed gas mass is expelled by AGN feedback leaving no
material available for forming stars.

Given that the collapsing gas flows inwards on a dynamical timescale,
$t_{\rm dyn}$, with a mass flow rate $\dot{M}$, the collapsing mass
can be written as $M = \dot{M}t_{\rm dyn} $. Then, using the fact that
the instantaneous mass flow rate is $\dot{M} = \sigma_{*}^{3}/(\alpha
G)$, the maximum mass of gas available for forming stars during each
collapse, will be
\begin{equation}\label{eq:mtotal}
M = \dot{M}t_{\rm dyn} \approx \frac{\sigma_{*}^{3}}{\alpha
  G}t_{\rm dyn} \approx 10^{10}\bigg(\frac{\sigma_{*}}{200 \,{\rm
    km\,s^{-1}}}\bigg)^{3}\bigg(\frac{t_{\rm dyn}}{10^{7} \,{\rm
    yr}}\bigg)\,{\rm M_\odot}.
\end{equation}
Since a fraction $\beta$ is assumed to be accreted by the black hole,
this leaves a maximum fraction of $1-\beta$ available for forming
stars, in the unlikely event that no material is removed due to
feedback.

However, we can estimate the mass of material removed by the AGN using
simple considerations. The total energy, $E$, injected by an AGN as a
result of accreting a mass $\beta M$, will be $E = \eta \beta M
c^{2}$. If this energy does work against the pressure, $P$, of the
ambient gas, it will inflate a bubble with volume, $V$, given
approximately by $E = 4PV$, assuming relativistic bubble contents. 
Using the definition of the gas pressure $P = \rho k_{\rm B}T/\mu m_{\rm p}$, where $\rho$ 
is the ambient gas density and $T$ is temperature, we can write $E = 4 M_{\rm dis}
k_{\rm B}T/\mu m_{\rm p}$, where $M_{\rm dis} \equiv \rho V$ is the
mass displaced in inflating the bubble. As previously noted,
\cite{popeetal} showed that a buoyant AGN-blown bubble will lift a
mass comparable to $M_{\rm dis}/2$ out from the central galaxy
 \footnote{It is difficult to estimate how high this material will be lifted because it depends 
strongly on the details of the bubble, the extra mass of material it is carrying, and the 
properties of the ambient atmosphere. \cite{popeetal} demonstrated that a bubble 
lifting additional material will rise to a height at which the average density of the bubble, 
plus the lifted mass, is equal to the ambient density. At this location the buoyancy force goes 
to zero, and the bubble cannot rise further unless it sheds some of the material. The greater 
the mass carried by the bubble, the less the bubble will rise. This effect directly limits the 
amount of energy extracted from the bubble. Because of this, it is not possible 
to say that the bubble will change the gravitational potential of the lifted mass by an 
amount defined by:  $E  = (M_{\rm dis}/2)\Delta \Phi$.}.

The mass of gas lifted out from the centre of a galaxy by AGN feedback can be related to
the mass of material that collapsed into the galaxy centre due to cooling-induced 
gravitational instability
\begin{equation} \label{eq:inout}
\frac{M_{\rm dis}/2}{M} \approx \eta \beta \frac{\mu m_{\rm p}
  c^{2}}{8 k_{\rm B}T} \approx 8
\bigg(\frac{\beta}{10^{-3}}\bigg)\bigg(\frac{T}{10^{7}\,{\rm
    K}}\bigg)^{-1}.
\end{equation}
According to equation (\ref{eq:inout}), an individual AGN outburst may
remove more gas than actually flowed into the galaxy centre. While
this may seem unphysical, it is not -- feedback could remove all of
the material that collapsed into galaxy \emph{and} additional ambient
gas. The fate of the gas lifted out of the
galaxy will depend on the AGN power and the depth of the external gravitational potential. 
There are two main possibilities: 1) if the outflow injects sufficient energy, material will be 
permanently expelled from the galaxy; 2) if not, the outflow will temporarily lift material 
out of the galaxy, later allowing it to fall back inwards. 

As shown in \cite{pope09}, AGN fuelled at a small fraction of the ambient 
cooling flow rate can power outflows that are capable of ejecting material from 
the potential of an elliptical galaxy. However, for the deeper gravitational potentials 
of galaxy groups and clusters (with virial temperatures greater than 1-2 keV) there 
are no black holes that are massive enough to sustain outflows that can expel gas 
from the potential. Despite this, typical AGN outflows in galaxy groups and clusters 
do still affect their environment by gently redistributing the gas within the gravitational 
potential.  Furthermore, any material that is lifted out of the centre of the host galaxy will 
eventually fall back inwards, thereby becoming part of a fountain-like flow.

Inspection of equation (\ref{eq:inout}) leads to the expectation that
AGN feedback should remove more ambient gas, per unit accreted mass, from cooler (lower
mass) systems. As described above, this material can be permanently
expelled from low mass elliptical galaxies. Thus, we conclude that
star formation in low mass elliptical galaxies would necessarily have
to occur in a limited window of opportunity: after the gravitationally
unstable material has collapsed into the centre of the galaxy, and
before it has been removed by AGN feedback. In other words, any
reservoir of material available for forming stars will be short-lived,
meaning that star formation episodes are comparatively rare in such
systems.

When applied to galaxy clusters, equation (\ref{eq:inout}) indicates that
the mass of material lifted out of the host galaxy by AGN feedback will 
approximately equal the mass of material which collapses into the galaxy due 
to gravitational instability. As a result, AGN feedback may also be able to shut-off
star formation in BCGs. However, the fountain-like flow described above would provide a
reservoir of material in and around the galaxy which is available for
forming stars. Consequently, is reasonable to expect some on-going star
formation in BCGs. Interestingly, this conclusion appears to be
compatible with observations; \cite{rafferty06} found BCG star
formation rates up to $\sim 100\,{\rm M_\odot \,yr^{-1}}$, which
corresponds to $10^{10}\,{\rm M_\odot}$ over $10^{8}\,{\rm yr}$.

Finally, considering the mass flow rates, we can show that the ratio
of black hole and stellar mass growth rates must be $\dot{M}_{\rm
  BH}/\dot{M}_{*} = \beta/(1-\beta) \approx \beta$. This suggests
that, as the age of the system becomes very large, the ratio of black
hole mass to bulge mass should also tend towards $\beta$. It is
encouraging to note that a value of $\beta \sim 10^{-3}$ agrees well
with observations \citep[e.g.][]{rix} \emph{and} is consistent with
our earlier constraints obtained by comparing the theoretical AGN duty
cycle with radio observations \citep[][]{best05,stas}.

\subsection{Possible implications for cosmological simulations}

The cooling-induced gravitational instability scenario for fuelling
AGN can only be captured by fluid simulations \emph{if} the following
are true: 1) the self-gravity of the gas is included in the
calculations, \emph{and} 2) the spatial resolution is fine enough to
track the evolution of the gravitationally unstable region. For a
collapsing mass $M$, with a velocity dispersion $\sigma_{*}$, the
region of importance has a size
\begin{equation}
R \sim \frac{G M}{\sigma_{*}^{2}} \approx 1
\bigg(\frac{M}{10^{10}\,{\rm
    M_\odot}}\bigg)\bigg(\frac{\sigma_{*}}{200\,{\rm
    km\,s^{-1}}}\bigg)^{-2}\,{\rm kpc}.
\end{equation}
Thus, for all plausible values of $M$ and $\sigma_{*}$, the collapsing
region is likely to be smaller than the resolved spatial scales in
cosmological fluid simulations. We note that the density of these
small, self-gravitating regions is likely to be high ($\sim
10^{-22}\,{\rm g\,cm^{-3}}$ for $M \sim 10^{10}\,{\rm M_\odot} $ and
$\sigma_{*} \sim 200\,{\rm km\,s^{-1}}$) and scales as $\rho \sim
M/R^{3} \propto \sigma_{*}^{6}/M^{2}$.

\section{Summary}
 
The aim of this article has been to explore the properties of AGN
feedback that is driven by the gravitational instability of hot gas in
the locality of a supermassive black hole. While it remains unclear
whether gravitational instability itself $\emph{is}$ responsible for
delivering fuel to AGN, we have shown that a mechanism which behaves
similarly could produce outcomes that are compatible with several key
observations of radio AGN in massive galaxies and clusters. The main
findings are summarised below:

\begin{enumerate}
\item Gas in the centre of a galaxy which periodically becomes
  gravitationally unstable provides a way of linking the AGN fuelling
  rate to galaxy's large scale environment without requiring gas to
  flow 10s of kiloparsecs before reaching the black hole.
\item According to this model, the AGN duty cycle scales as $\delta
  \propto L_{\rm X}/\sigma_{*}^{3}$, where $L_{\rm X}$ is the X-ray
  luminosity of the hot gas in a cluster and $\sigma_{*}$ is the
  stellar velocity dispersion at the centre of the galaxy which hosts
  the AGN.
\item Applying simple scaling relations to this model, we find that
  the duty cycle of radio AGN in massive galaxies should scale as
  $\delta \propto \sigma_{*}^{6} \propto M_{\rm BH}^{1.5}$, in
  reasonable agreement with observations.
\item The region which collapses may be very small ($< 1\,{\rm kpc}$)
  and is, therefore, difficult to capture directly in cosmological
  numerical simulations.
\item The model predicts that the typical AGN outburst duration should
  scale as $t_{\rm dyn} = \delta t_{\rm age}/N \propto L_{\rm
    X}/(\sigma_{*}^{3}N)$, where $t_{\rm age}$ is the age of the
  cluster and $N$ is the number of outbursts during this
  time. Plausible values for $t_{\rm age}$ and $N$ imply that radio
  AGN outbursts fuelled by gravitational collapse can extend up to
  $\sim 10^{7}-10^{8}\,{\rm yrs}$, which is consistent with
  observations of AGN-blown bubbles.
\item Using simple arguments we have shown that AGN feedback is likely
  to remove more gas, per unit accreted mass, from the centre of a 
  lower mass galaxy than a BCG. By completely expelling material from lower mass
  elliptical galaxies, AGN feedback can dramatically reduce their star
  formation rates. In contrast, AGN feedback cannot completely expel
  gas from a galaxy cluster. As a result, there may be more material
  in and around BCGs which is available for forming stars. Thus, it
  may be more difficult to completely shut-off star formation in BCGs
  than in lower mass galaxies.
\end{enumerate}

\section{Acknowledgements}
We thank the anonymous referee for helpful comments that improved this work. 
ECDP would like to thank CITA for funding through a National
Fellowship. JTM acknowledges financial support from the Canadian
National Science and Engineering Research Council (NSERC). SSS
acknowledges the Australian Research Council (ARC) for a Super Science
Fellowship.

\bibliography{database} \bibliographystyle{mn2e}

\label{lastpage}

\end{document}